\documentclass[pra,twocolumn,showpacs,superscriptaddress,floatfix,nofootinbib]{revtex4-2}

\usepackage{color}
\usepackage{hyperref}  
\usepackage{graphicx}
\usepackage{amsfonts}
\usepackage{footmisc}

\usepackage{fancyhdr}
\let\a=\alpha   \let\g=\gamma  \let\d=\delta \let\e=\varepsilon
  \let\h=\eta    \let\k=\kappa \let\l=\lambda
\let\m=\mu    \let\n=\nu         \let\p=\pi    \let\r=\rho
 \let\t=\tau    

 \let\D=\Delta  \let\L=\Lambda 
         
\let\O=\Omega 

\font\tenmib=cmmib10\font\sevenmib=cmmib7\font\fivemib=cmmib5%
\mathchardef\Bl   = "0515  

\def\Bl   {{\mbox{\boldmath$ \lambda$}}}

\def\Bf   {{\mbox{\boldmath$ \phi$}}}

\def\BDpr {{\mbox{\boldmath$ \partial$}}}

\def\eqalign#1{\null\,\vcenter{\openup\jot
  \ialign{\strut\hfil$\displaystyle{##}$&$\displaystyle{{}##}$\hfil
      \crcr#1\crcr}}\,}

\def\CC{{\mathcal C}}
\def\EE{{\mathcal E}}\def\DD{{\mathcal D}}
\def\NN{{\mathcal N}}

\def\GG{{\mathcal G}}

\def\uu{{\V u}}\def\kk{{\V k}}\def\xx{{\V x}}\def\ff{{\V f}}

\def\hh{{\V h}}\def\qq{{\V q}}\def\pp{{\V p}}
\def\T#1{{#1_{\kern-3pt\lower7pt\hbox{$\widetilde{}$}}\kern3pt}}

\def\nn{{\bf n}}\def\vv{{\bf v}}

\def\ie{{\it i.e.\ }}
\def\dpr{{\partial}}
\def\defi{{\buildrel def\over=}}

\def\otto{\,{\kern-1.truept\leftarrow\kern-5.truept\to\kern-1.truept}\,}
\def\Pprod{\prod^{\kern-1mm\raise.0mm\hbox{$\leftarrow$}}}
  
\newdimen\xshift \newdimen\xwidth \newdimen\yshift \newdimen\ywidth

\def\ins#1#2#3{\vbox to0pt{\kern-#2pt\hbox{\kern#1pt #3}\vss}\nointerlineskip}

\def\eqfig#1#2#3#4#5{
\par\xwidth=#1pt \xshift=\hsize \advance\xshift
by-\xwidth \divide\xshift by 2
\yshift=#2pt \divide\yshift by 2
{\hglue\xshift \vbox to #2pt{\vfil
#3 \includegraphics{#4.eps}
}\hfill\raise\yshift\hbox{#5}}}

\def\V#1{{\bf #1}}
\def\lis#1{{\overline#1}}

\def\tende#1{\,\vtop{\ialign{##\crcr\rightarrowfill\crcr
 \noalign{\kern-1pt\nointerlineskip} \hskip3.pt${\scriptstyle
   #1}$\hskip3.pt\crcr}}\,}
\def\eg{{\it e.g.\ }}

\def\0{\noindent}
\def\*{\vskip2mm}
\def\media#1{\langle #1 \rangle}
\def\Eq#1{\label{#1}}
\def\equ#1{(\ref{#1})}

\font\titolo=cmbx12%
\def\iniz{\setcounter{equation}{0}}
\def\be{\begin{equation}}\def\ee{\end{equation}}

\newcounter{appendice}

\def\alert#1{{\color{ired}#1}}
\definecolor{iblue}{RGB}{65,105,225}
\definecolor{ired}{RGB}{220,20,60}
\definecolor{igreen}{RGB}{50,205,50}
\definecolor{ipurple}{RGB}{75,0,130}
\definecolor{iochre}{RGB}{218,165,32}
\definecolor{iteal}{RGB}{51,204,204} 
\definecolor{imauve}{RGB}{204,51,153}

\def\ap{{\it a priori }}
\iniz

\begin{document}

\alert{\centerline{\titolo NS and equivalence conjectures.}}
\vskip1mm
\centerline{\bf Giovanni Gallavotti}
\centerline{Universit\`a ``La Sapienza'' and INFN, Roma}
\centerline{\today}

{\vskip3mm}
\noindent {\bf Abstract}: {\it General considerations on the
Equivalence conjectures and a review of few mathematical results.}{\vskip3mm}

\def\SEC{Introduction}
\section{\SEC}
\label{sec1}
\def\Dot#1{{\bf\dot#1}}

Reversible equations conjectured to be equivalent for the purpose
of modeling
stationary states, under large scale forcing, of incompressible
Navier-Stokes evolutions have been introduced since the
90's, \cite{SJ993,Ga997}. Equivalence has been first conjectured to
be asymptotic at vanishing viscosity, then at all viscosity for
observables of large scale, \cite{Ga020b}, or just for
observables of scale up to $\sim$Kolmogorov's scale, \cite{MBCGL022}.

There are very few rigorous results, stressed
in \cite{Ga020b,MBCGL022}, supporting the conjectures. And some
breakthrough results, \cite{Ru982,Li984}, on the classical NS
equation can be directly applied to the reversible equations.

The NS equations, incompressible and in a periodic container
$\O=[-\p,\p]^d$, $d=2,3$,
deal with a velocity field which can be expressed in terms of its
Fourier's coefficients as
\be \uu(\xx)=\sum_{\V0\ne\kk\in Z^d,\atop c=1,\ldots,d-1} u^c_\kk 
{\V e^c_\kk}
e^{-i \kk\cdot\xx}
\Eq{e1.1}\ee
where ${\V e}^c_\kk$, $c=1,\ldots,d-1$, are $d-1$ unit ``elicity'' vectors,
orthogonal to $\kk$ and with ${\V e}^c_{-\kk}=-{\V e}^c_\kk$, and
$u^c_{-\kk}=\lis u^c_{\kk}$ are the complex harmonics of $\uu$,
with $|u^c_{\V0}|\equiv0$ (fixing the baricenter).

With $\uu$ so represented the well known NS equations are
$\BDpr_\xx\cdot\uu(\xx)=0$ (incompressibility) and:

\be \dot \uu(\xx)=-
(\T \uu(\xx)\cdot\T\BDpr_{\raise 2mm\hbox{$\scriptstyle\xx$}})
\uu(\xx)+\n \D\uu(\xx)-\BDpr_\xx
p(\xx)+\V f(\xx)\Eq{e1.2}\ee
with $\n$= kinematic viscosity. Or, in terms of the $u^c_\kk$:

\be \eqalign{
&\dot u^c_\kk=\sum_{\kk_1+\kk_2=\kk\atop a,b}
T^{a,b,c}_{\kk_1,\kk_2,\kk}
u^a_{\kk_1}u^b_{\kk_2} -\n \kk^2 u^c_\kk +f^c_\kk\cr
&T^{a,b,c}_{\kk_1,\kk_2,\kk}=- ({\V e}^a_{\kk_1}\cdot\kk_2)\,
({\V e}^b_{\kk_2}\cdot{\V e}^c_{\kk}),\quad \kk=\kk_1+\kk_2\Eq{e1.3}\cr}\ee
with $\kk^2\defi\sum_{i=1,\ldots,d}k_i^2$ and the forcing $\V
f\ne \V0$ is supposed fixed once and for all and to act only on
'large scale': $f^c_\kk=0$ unless
$0<|\kk|=\max_{i=1,\ldots,d}|k_i|\le k_f<\infty$.

Without restriction, suppose $||\V f||^2=\sum_{\kk,c}
|f^c_\kk|^2=1$: so the equation has viscosity as the only
free parameter, whose inverse will also be called Reynolds number
$R=\n^{-1}$. The equation will be called ``INS'', or {\it irreversible} NS.

\def\SEC{Equivalent equation}
\section{\SEC}
\label{sec2}
\iniz
\def\Dot#1{{\bf\dot#1}}

Equivalent equations can be obtained by replacing the viscous
force $-\n \kk^2 u^c_\kk$ by $-\a(\uu) \kk^2 u^c_\kk$ determining
$\a$ so that the equation:

\be \dot u^c_\kk=\sum_{\kk_1+\kk_2=\kk\atop a,b}
T^{a,b,c}_{\kk_1,\kk_2,\kk}
u^a_{\kk_1}u^b_{\kk_2} -\a(\uu) \kk^2 u^c_\kk
+f^c_\kk\Eq{e2.1}\ee
will admit a selected observable as an
exact constant of motion.

In \cite{Ga997,Ga020b} the selected observable is the
``Enstrophy'':

\be\eqalign{\DD(\uu)&=\sum_{c,\kk} \kk^2
|u^c_\kk|^2
=\int \DD(\uu,x)\,\frac{dx}{(2\p)^d}\cr
\DD(\uu,x)&=\frac12\sum_{i,j}(\dpr_i u(x)_j+\dpr_j
u(x)_i)^2 \cr
} \Eq{e2.2}\ee
Other
observables have been considered (\eg in \cite{SDNKT018} the
selected observable is $E=\sum_{c,\kk} |u^c_\kk|^2$). Selecting
$\DD$ leads, if $d\ge3$, to:

\be \eqalign{
\a(\uu)=&\frac{\sum^*_{\kk_1,\kk_2\kk_3}\sum_{a,b,c} \kk_3^2
u^a_{\kk_1} u^b_{\kk_2} \lis u^c_{\kk_3}
T^{a,b,c}_{\kk_1,\kk_2\kk_3}}{\sum_{c,\kk}\kk^4 |u^c_\kk|^2}
\cr
 &+\frac{\sum_{c,\kk} f^c_\kk \,\kk^2\,\lis
u^c_\kk}{\sum_{c,\kk}\kk^4 |u^c_\kk|^2}\cr}\Eq{e2.3}\ee
where the $*$ reminds $\kk_1+\kk_2+\kk_3=\V0$. See Eq.\equ{e4.2}
for a possibly more natural expression of $\a(\uu)$.

If $d=2$ the multiplier $\a$ would simply be the second term in
Eq.\equ{e2.3}: because the first term would
cancel.\footnote{\small By the well known
identity which implies the Enstrophy conservation (only) in the
$2$-dimensional Euler-equation.\label{ens}} Furthermore
selecting $E$, instead of $D$, yields $\a=\frac{\sum_{c,\kk} f^c_\kk \,\lis
u^c_\kk}{\sum_{c,\kk}\kk^2 |u^c_\kk|^2}$ in any
dimension \footnote{\small By
the identity which implies the energy conservation in the
Euler equation.\label{ene}}

The Eq.\equ{e2.1} will be called RNS, {\it reversible \rm NS}: because if
$t\to \uu(t)$ is a solution for Eq.\equ{e2.1} also $-\uu(-t)$ is
a solution. Correspondingly $\a(\uu)$ will also be named ``{\it
reversible viscosity}''.

Here only properties of the RNS equation with $\a$ such that the
enstrophy is constant will be considered.

The equivalence conjectures concern the stationary distributions
of INS and of RNS and the averages that they assign to the
``{\it local observables}'' $O(\uu)$, which are functions $O(\uu)$ of the
velocity fields which depend on finitely many harmonics
$u^c_\kk$, possibly subject to the further condition that the
waves $\kk$ are $|\kk| \ll K_\n$ where $K_\n$ is Kolmogorov's
inverse length scale $K_\n=(\frac{En}{\n^2})^{\frac14}$.

To formulate mathematically precise conjectures introduce the
{\it regularized equations} $INS^N$ and $RNS^N$ with ultraviolet
cut-off $N$:
\*
\0{\bf Definition:} {\it The equations $INS^N$ and $RNS^N$ are
the Eq.\equ{e1.3},\equ{e2.1} 
with $\n$ and $\a(\uu)$ respectively as above with the further
restriction on the $u^c_\kk$ that $|u^c_\kk|=0, |\kk|>N$.}
\*
Therefore both regularized equations are ODE's on a phase space
of (real) dimension $M=(d-1)((2N+1)^d -1)$ as each component
has $0<|\kk|=\max_i |k_i|\le N$.

Hence it makes sense to consider initial data in $R^M$ randomly
selected with a distribution $\r(\uu)d\uu$, with $\r$ a
continuous density \ie ``volume continuous''. Starting evolution
$t\to S_t(\uu)$ with an initial datum $\uu$ so chosen, it is
assumed:

\*
\0{\bf Hypothesis:} {\it Given $\n$ or $D$ there are, for the
$INS^N$ or $RNS^N$ evolutions, finitely many stationary ergodic
probability distributions denoted, for $i=1,\ldots,\NN$,
$\m^N_{\n,i}$ or, respectively, $\g^N_{D,i}$ which control the
statistics of the local observables $O$. This means that, on motions starting
with initial data chosen with a volume-continuous distribution,
the average of such observables is given, with probability $1$,
by $\m^N_{\n,i}(O)$ or, respectively, $\g^N_{D,i}(O)$ for some $i$ .}
\*

When the motion is ``chaotic'', \cite{Ru995,Ru977,Ru989}, (or, a
particular case, satifies the ``Chaotic
Hypothesis'', \cite{GC995,GC995,Ga013b}) and has $\NN$ attractors
then each $\m_i$ is called a ``SRB-measure'', \cite{Ru989}.

It is expected that in most cases $\NN=1$, {\it greatly
simplifying} the hypothesis: which embodies the classical ergodic
hypothesis if applied to the chaotic microscopic motions of
Hamiltonian systems of many particles.
\*

{\bf Remark:} the above hypothesis is intended to apply also to cases
in which the attractors are periodic orbits (typically if
viscosity is large). In \cite{Ga020b} it is suggested that the
reason for its validity at fixed $\n$ and large enough $N$ could
be looked in the microscopic motions, from which the NS evolution
is derived as a scaling limit without modifications to the
microscopic equations. But at fixed cut-off $N$ the hypothesis
can be related, and possibly hold, to the chaoticity of the motions 
only at small enough $\n$: manifestly a less interesting case.
\*

However the latter key point will not be further discussed here,
as attention is devoted to mathematical properties of the
regularized equations in the light of the following conjectures
and the regularization removal, \ie $N\to\infty$.

\def\SEC{Equivalence conjectures}
\section{\SEC}
\label{sec3}
\iniz
Given the UV cut-off $N$ the stationary SRB distributions (think,
at first, that there is only one such) form a collection $\EE^N$
of probability ditributions on $M$, the 
$(d-1)((2N+1)^d-1)$-dimensional  phase space,
parameterized by $\n$ in the $INS^N$ case or by $D$ in the
$RNS^N$ case and possibly other $\NN$ labels distinguishing the
SRB distributions on the $\NN$ attractors (there is no relation
between the cut-offf $N$ and the number of degrees of freedom $\NN$).

Denote $\EE^N$ and $\GG^N$ the collection of the SRB
distributions $\m^N_\n, \n>0$ for $INS^N$, or repectively,
$\g^N_D, D>0$ for $RNS^N$: such collections will be called {\it
viscosity ensemble} or {\it enstrophy ensemble}.
\*

\0{\bf Conjecture 1:} {\it (1) if $\n$ is small enough the number $\NN$ of
attractors for $INS^N$ with viscosity $\n$ and average enstrophy
$\m^N_\n(\DD)=D$ is eventually (as $N\to\infty$) the same as the number of
attractors for $RNS^N$ with enstrophy $D$.\footnote{\small In the
relation $\m^N_\n(\DD)=D$ given $\n$ the average $D$ depends also
on $N$: the extra label $N$ on $D$ will be always omitted in the
following to simplify the notation if clear from the context.}\\
(2) If $O$ is an observable and $\m^N_\n(\DD)=D$ it is
\be\lim_{\n\to0} \m^N_\n(O)=\lim_{D\to\infty} \g^N_{D}(O)\Eq{e3.1}\ee
}
\*
The conjecture, \cite{Ga997,Ga013,Ga019c}, can be regarded as a
``homogeneization'' property: at large $D$ the $RNS^N$ equation
generates chaotic motion (a feature shared quite generally by
strongly forced ODE's) and $\a(\uu)$ fluctuates around a constant
value and induces averages of observables equal to those of
$INS^N$ with constant viscosity, as in \cite{GL014}.  A second conjecture is:
\*

\0{\bf Conjecture 2:} {\it (1) if $N$ is large enough the number $\NN$ of
attractors for $INS^N$ with viscosity $\n$ and average enstrophy
$\m^N_\n(\DD)=D$ is the same as the number of attractors for
$RNS^N$ with enstrophy $D$.\\
(2) If $O$ is a local observable and $\m^N_\n(\DD)=D$ it is
\be\lim_{N\to\infty} \m^N_\n(O)=\lim_{N\to\infty} \g^N_{D}(O)\Eq{e3.2}\ee
}
\*

Therefore conjecture 1 deals with the limit $\n\to0$ at fixed
UV-cut-off $N$ and holds for any observable, while
conjecture 2 deals with the physically relevant limit
$N\to\infty$ at fixed $\n$ and holds for local (\ie large scale) observables.
A much weaker conjecture:
\*
\0{\bf Conjecture 3:} {\it (1) same as in conjecture 2.\\
(2) If $O$ is an observable localized on scale sufficiently small
compared to Kolmogorov's scale $K_\n=(\frac{\n
D}{\n^3})^{\frac14}$ and $\m^N_\n(\DD)=D$ it is
\be\lim_{N\to\infty} \m^N_\n(O)=\lim_{N\to\infty} \g^N_{D}(O)\Eq{e3.3}\ee
}
\*
This restricts the observables $O$ to depend on the Fourier's
components $u^c_\kk$ of the velocity with $|\kk|<K_\n$, provided
$\m^N_\n(\DD)=D$ as in the previous conjectures. Hence an
observable $O$ depending on just one $\kk$ will be a local
observable relevant for conjecture 2 but not for conjecture 3
unless $|\kk|$ is ``sufficiently smaller'' than $K_\n$.

Conjecture 3 is introduced in \cite{MBCGL022} to cover at least
the results of the corresponding simulations: the simulations
were not developed enough to allow stating that the failure of
conjecture 2 on observables of scales over $\simeq\frac18 K_\n$,
as apparently shewed by the simulations, could be firmly confirmed;
the point was left for consideration in future work.
\*

{\bf Remark:} Introducing the ``viscosity'' and ``enstrophy''
ensembles leads to a strong analogy between equilibrium
statistical mechanics (where the finite volume regularizes the
dynamics) and stationary properties of NS evolution (where the
UV-cut-off regularizes the dynamics): aspects of the analogy have
been pointed out, for instance, in
\cite{Ru012,Ru014,Ga020b}.
The conjectures (in particular conjecture 2) make the
thermodynamic limit (infinite volume) analogous to the $N\to\infty$ limit.

\def\SEC{$RNS^N$-uniform regularity}
\section{\SEC}
\label{sec4}
\iniz

It is well known that in dimension $3$ an algorithm for the
construction of a smooth solution for INS with $\CC^\infty$-smooth
initial data and smooth force is an open
problem, \cite{Fe000,BV019}: the difficulty being to establish
an \ap bound on the enstrophy, \cite{CKN982}.
 
In the $RNS^N$ case the enstrophy of a smooth datum $\uu$ is finite and
evolves at time $t$ into $\uu(t)=S^N_t\uu$ with the same enstrophy. To
investigate the regularity it is natural to study first the size
of $\a(u)$ in a velocity field of given enstrophy $D$.
\*
\0{\bf Theorem 1:} {\it If $\uu\in \CC^{\infty}$ has enstrophy $\DD(\uu)=D$
then the multiplier $\a(\uu)$ in $RNS^N$, Eq.\equ{e2.3} is
bounded by
\be |\a(\uu)|\le C_1 (D^{\frac12}+D^{-\frac12})\Eq{e4.1}\ee
with $C_1$ a universal constant, independent of the UV-cut-off
$N$. \cite{MBCGL022}.}
\*

This kinematic inequality (\ie depending on $\uu\in \CC^{\infty}$
and unrelated to the $RNS^N$) is obtained by combining the
H\"older and Sobolev's inequalities, see for
instance \cite[Appendix A]{MBCGL022}, applied to
$\L(\uu)=-\int [(\T \uu\cdot\T\BDpr) \uu]\cdot\D\uu\,d\xx$ which
appears in the expression of the first addend in Eq.\equ{e2.3}
rewritten in the form:
\be \a(\uu)=\frac{\L(\uu)+\int \ff\cdot\D\uu \,d\xx}{\int
(\D\uu)^2\,d\xx} \Eq{e4.2}\ee

A remarkable regularity, uniform in time and in the UV-cut-off,
holds for solutions of the $RNS^N$.

Suppose $0<\e <\a(\uu(t))\le \k $, for some $\e,\k>0$,
and that the initial data $\uu(0)$ and the forcing $\ff$ satisfy
$|\uu_{\kk}|,|\ff_\kk|< c_p |\kk|^{-p}$ for all $p>0$ (recall
that we consider only initial data and force with a finite
number of modes, $\le N$, for simplicity). Let
$a(t,\t)=\int_\t^t
\a(\uu(t')dt'$; then $\e (t-\t)<a(t,\t)<\k (t-\t)$. 

\*
\0{\bf Theorem 2:} {\it If $\uu(t)$ is a solution of $RNS^N$ with
enstrophy $D$ and $\a(\uu(t))>\e>0$ then $\uu(t)$ is 
$\CC^\infty$-regular with $\CC^k$ norm $||\uu(t)||_{\CC^k}<
c_k(\e,||\uu(0)||_{\CC^\infty})$ where  $c_k$ is independent of $t$ and
of the UV-cut-off $N$. \cite{MBCGL022}.}
\*

\0{\it proof:} Following, for instance, \cite{Ga002} and clarifying the notations, 
write $\uu_\kk(t)=e^{-a(t,0) \kk^2}\uu_\kk(0)+\int_0^t
e^{-a(t,\t)k^2}( {\bf n}_\kk(\uu(\t))+\ff_\kk)d\t$, where ${\bf
n}_\kk(\uu(\t))$ is the non-linear term of the NSE.  Therefore,
the sum of the first and last term can be bounded by $\frac{c_p
}{|\kk|^p}$, $c_p=||\D^p\uu(0)||_2+||\D^p\ff||_2)$, while the
integral is bounded by
\be\int_0^t e^{-\e
k^2(t-\t)}||\BDpr\T\uu(\t)||_2 ||\uu(\t)||_2 \,d\t 
\le \frac{\sqrt{D \EE }}{\e \kk^2}\Eq{e4.3}\ee
where $\EE$ is an \ap bound on $\sum_\kk|\uu_\kk|^2$ (\eg
$D$ itself as $|\kk|\ge1$)
so that adding the two bounds:
$|\uu_\kk(t)|^2< \frac{C_2}{\kk^2}$ for a suitable $C_2$ (\eg
$C_2=\frac{\sqrt{D\EE}}\e+c_2$).  Therefore, again,
$|\uu_\kk(t)|$ can be bounded by adding $\frac{c_p}{|\kk|^p}$,
contributed from the initial datum, and a bound on
		
\be \eqalign{&\int_0^t e^{-\e \kk^2(t-\t)}
\sum_{\kk_1+\kk_2=\kk}
\frac{|\kk_1||\uu_{\kk_1}|\,
|\kk_2|^2|\uu_{\kk_2}|}{|\kk_1|\,|\kk_2|} \,d\t
\cr}\Eq{e4.4}
\ee
A bound on the latter integral is obtained via the Schwartz
inequality and the remark that $\kk_1+\kk_2=\kk$ implies
$|\kk_1|\,|\kk_2|\ge \frac{k_0}2 |\kk|, k_0=1$, and
\be\eqalign{
&\sum_{\kk_1+\kk_2=\kk}\kern-3mm
\frac{{|\kk_1|}|\uu_{\kk_1}| |\kk_2|^2|\uu_{\kk_2}|}{|\kk_1||\kk_2|}
\le 2 C_2 \sum_{\kk_1+\kk_2=\kk}\kern-3mm
\frac{|\kk_1|\,|\uu_{\kk_1}|}{|\kk_1||\kk_2|}\cr
&\le 2 C_2 \sqrt{D} \left(\sum_{\kk_1+\kk_2=\kk}
\frac1{(|\kk_1||\kk_2|)^2}\right)^{\frac12}
\cr
&\le 2^{1+\frac18} C_2 |\kk|^{-\frac18} \sqrt{D}
\left(\sum_{\kk_1+\kk_2=\kk}\frac1{(|\kk_1||\kk_2|)^{2-\frac14}}\right)^{\frac12}
\cr
&\le 2^{1+\frac18} C_2 \sqrt{D}|\kk|^{-\frac18} 
\left(\sum_\nn \frac1{|\nn|^{4-\frac12}}\right)^{\frac12}\cr}\Eq{e4.5}\ee
where $\kk_1$ has been changed to $\nn$ just to make clear that summing
over $\kk_1+\kk_2=\kk$ allows using the Schwartz inequality. Hence
integration over $t$, as in Eq.\equ{e4.3}, yields for suitable
$C_3$, proportional to $\sqrt D$:
\be|\uu_\kk(t)| \le \frac{C_3}{k^{2+\frac18}}\Eq{e4.6}\ee
\hglue.2cm Thus if $D$ is finite the bound $|\uu_\kk|<\g k^{-2}$, Eq.~\equ{e4.3},
can be improved into $|\uu_\kk|<\g_1 |\kk|^{-2-\frac18}$.

Iterating a {\it autoregularization} phenomenon sets in and
\be ||\uu_\kk(t)||_2 \le \frac{\g_p}{k^{2+\frac14 p}}\qquad {\rm for\ all}
\ p\ge1\Eq{e4.7}\ee
so that $\uu(t)$ is a $C^\infty$-functions and all its derivatives can be
bounded in terms of the enstrophy $D$, uniformly in $N$. See Sec. 3.3 in
\cite{Ga002} for related results on the classic autoregularization.
\*

\0{\bf Remarks:} (1) the theorem shows that the condition $\a(\uu(t))>\e$ for
some $\e>0$ has an {\it extremely unlikely possibility}. Besides
providing a well defined prescription to construct uniformly
smooth approximations of
the RNS equations as sequences of solutions to $RNS^N$,
it would imply that the $RNS^N$ attractors consist of uniformly $\CC^\infty$
velocities (\ie with $\CC^k$ norms uniformly bounded for each $k$ and
independently of $N$).
\\
(2) the simulations with large $N$ show that $\a(\uu(t))$ is
observed, after a transient, not only $>0$ but also quite close to $1$:
in \cite{MBCGL022} evidence is provided that this might be an
illusion: following the evolution of $\a(t)$ on typical $RNS^N$
solutions it is found that as $N$ increases the negative values
of $\a$ have a rapidly decreasing probability. So the negative
values of $\a$ might be not observable within the precision of
the simulation and the time available to it.
\\
(3) of course the latter comment also indicates that, if the
above conjectures are confirmed, there could be an alternative
way of studying the INS equation regularity: rather than looking
for solutions in suitable function spaces it would be
particularly relevant to study solutions of the $RNS$ equations
trading the search of singularities with the search of extremely
unlikely events wit $\a<0$.
\\
(4) the closeness to $1$ of $\a(\uu(t)$ in the simulations
mentioned in (2) above, raises questions on whether the
restriction on the notion of locality in conjecture 3 can be
considered as really needed: it arises from simulations in which
not only $\a(\uu(t))$ stays $>0$ but also fluctuates close to
$1$. If this is not due to the precision of the simulation and
the time available to it, it is difficult to believe that the
difficulties (which seem unsurmountable at constant viscosity)
disappear if viscosity only slightly fluctuates, leaving valid
the key \ap bounds based on the positivity of the viscosity in,
for instance, \cite{CKN982}.

\def\SEC{INS Lyapunov spectrum}
\section{\SEC}
\label{sec5}
\iniz

The linearization of the $INS^N$ or $RNS^N$ flows is the $M
\times M$ matrix, $M=(d-1)((2N+1)^3-1)$, formally defined as
$J_{c,\kk;b,\hh}= \frac{\dpr\dot u^c_\kk}{\dpr u^b_\hh}$. To
really define it the $u^c_\kk$ can be represented a real $M$
components vectors $\{U_s\}_{s=0,\ldots,M}$ holding the
$u^c_\kk$, if $d=3$, as
\\
{\phantom.}\kern2mm (a) for $c=0$: $U_{2 i}$ are real parts of
$u^0_\kk$ after labeling half of the $\kk$ arbitrarily with
$i\in[0,M/4)$; and $U_{2 i+1}$ are the corresponding imaginary
parts of $u^0_{\kk}$
\\
{\phantom.}\kern2mm (b) for $c=1$: $U_{2 i+M/2}$ and $U_{2i+1+M/2}$ are labeled,
from the $u^1_\kk$, likewise as $i\in [0,M/4)$.

Consider first the $INS^N$ equations.

Then the equation can be written $\dot U_s=N_s(U,U)$ and its
Jacobian as $J_{s;r}(\uu)=\frac{\dpr \dot U_s}{\dpr
U_r}$.\footnote{A (arbitrary) way to define, in $d=3$, the labels $i$ is to
consider first the $\kk=(k_0,k_1,k_2)$ with $k_0>0,k_1=0,k_2=0$
assigning them the labels $i=k_0-1\in [0,N-1]$, then consider the
$\kk=(k_0,k_1>0,k_2=0)$ assigning them $i=N+(k_1-1)(2N+1)+k_0+N$
(hence $0\le i<2N(N+1)=((2N+1)^2-1)/2$), and finally assign to
$\kk=(k_0,k_1,k_2>0)$ the label
$i=2N(N+1)+(k_2-1)(2N+1)^2+(k_1+N)(2N+1)+k_0+N$ and $0\le i<
((2N+1)^3-1)/2)$. The total number of labels $i$ is
$n=((2N+1)^3-1)/2)$. For each $\kk$ there are two complex
components $u^c_\kk, c=0,1$: then given $\kk,c$ assign
$U_{2i+n\,c}=Re(u^c_(\kk)), U_{2i+1+n\,c}=Im(u^c_(\kk))$.}

The matrix $J$ can also be represented as an operator on the
velocity fields acting by multiplication by
$T_{c;d}(\xx)=-\dpr_{d}\dot u_c(\xx)$, \ie $(T\vv)(\xx)_c=\sum_d
T_{c;d}(\xx)\vv_d(\xx)$, to which the viscosity contribution has
to be added.

The symmetrized $J(\uu)$, if $W_{c,d}(x)= -\frac12 (\dpr_{d}u_c(\xx)+
\dpr_{c}u_d(\xx)) $ is therefore:
\be J^s_{c,d}(\uu)= \n \d_{cd}\D +W_{c,d}(x) \Eq{e5.1}\ee
Following \cite{Ru982} introduce $w(\xx)$ as the largest
eigenvalue of the matrix $J^s(\xx)$. The (negative of the)
Schr\"odinger operator $H=\n \D+w$ will be considered as an
operator on $L^N_2(\O)$ of divergenceless velocity fields. The
$w(\xx)$ could be bounded by $(Tr W^2)^{\frac12}$ which is
bounded above by the enstrophy density $\frac12\sum_{a,b}
(\dpr_{a}u_b(\xx))^2=\e(\xx)$; the bound, \cite{Ru982}, can be
improved into:
\be w(\xx)\le \frac{(d-1)}{d}\e(\xx)\Eq{e5.2}\ee
as shown in \cite{Li984} taking advantage that the trace of
$W(\xx)$ is zero.

Denoting $\m^{0,N}(\uu(0))\ge \ldots\ge\m^{M,N}(\uu(0))$ the Lyapunov exponents
for an ergodic stationary distribution $\r$ for the $IRS^N$ or $RNS^N$, the sum of the 
$n$ largest exponents, defined for $\r$-almost all initial $\uu(0)$,  is
\be\kern-3mm\eqalign{
\sum_{i=0}^{n-1}& \m^{i,N}(\uu(0))\cr&=\lim_{t\to\infty}\frac1{2t}\log
||\Bf_0(t)\wedge \Bf_1(t)\wedge \ldots\wedge \Bf_{n-1}(t)||^2\cr}\Eq{e5.3}\ee
with $\Bf_j(t)=S^N_t(\Bf(0))$ for almost all choices of the $n$
fields $\Bf_j(0)$'s in the $M$-dimensional phase space.

The time derivative of the $\log$ in Eq.\equ{e5.3} yields the expectation value
of $J(\uu(t))$ in the state
$\frac{\Bf_0(t)\wedge\ldots\wedge \Bf_{n-1}(t)}
{||\Bf_0(t)\wedge\ldots\wedge \Bf_{n-1}(t)||}$ and
via the max-min principle leads to an estimate about the Lyapunov
exponents in terms of the the eigenvalues $a^{k,N}(\uu(t))$, in
decreasing order, of the operator $\n\D+w$ on $L_2(\O)$, \cite[p.291]{Ru982}:

\*
\0{\bf Theorem 3:} {\it For all $n$:
\be\sum_{k=0}^{n-1}\m^{(k)}\le \sum_{k=0}^{n-1}\media{a^{k,N}}
\le \sum_{k=0}^{n-1}\media{a^{k,\infty}}\Eq{e5.4}\ee
where the $\media{\cdot}$ denote time average or, equivalently,
average with respect to the invariant $\r$.}
\*
This is obtained in \cite[Eq.(1.7)]{Ru982} for the INS evolution,
and as a consequence of the variational principle the {\it
argument applies as well to $INS^N$} (and to $RNS^N$ if
$\a(\uu(t))$ is
eventually $>\e>0$ for some $\e$).

A particularly remarkable estimate is derived, \cite{Ru982,Li984}, as
\*
\0{\bf Theorem 5:} {\it For $d=2,3$ and $\g\ge0$
\be \sum_{a^{k,N}\ge0} (a^{k,N})^\g\le
L_{\g,d} \n^{-\frac{d}2}\int_\O \DD(\uu,x)^{\frac\g2+\frac{d}4} dx\Eq{e5.5}\ee
with $L_{\g,d}<\infty$ for $\g>0$ if $d=2,3$ and
$L_{0,3}<\infty$. Furthermore the best constant $L_{\g,d}$ is
$\O$--independent.}
\*

\0(a) Whether $L_{0,2}<\infty$ is an open problem. The restriction
$\g\ge1$ in  \cite{Ru982} is improved to $\g\ge0$ in \cite{Li984}
if $d=3$ and $\g>0$ if $d=2$.
\\
(b) The interest of the case $\g=0$ is that it estimates the
number $\lis\NN$ of positive Lyapunov exponents as bounded in
terms of the viscosity and of $\h$=average energy dissipation per
unit time (finite for all $N$ in $INS^N,RNS^N$, and for $INS$
(\ie $N=\infty$) conjectured to
be finite and to have a positive finite limit even as
$\n\to0$, \cite{Ta935,Kr975a}).
\\
(c) Hence for $\g=0,d=3$ the bound on $\lis\NN$, implied by
H\"older's inequality with $p=\frac43,q=4$ applied to
Eq.\equ{e5.5}, is (using also convexity of $x\to x^{\frac34}$):
\be \lis \NN\le L_{0,3}\,\frac{|\O|}{3^{\frac14}}\,
\Big(\frac{\n\media{\DD(\cdot)}}{\n^3}\Big)^{\frac34}
=\frac{L_{0,3}|\O|}{3^{\frac14}}\,
\Big(\frac{\h}{\n^3}\Big)^{\frac34}\Eq{e5.6}\ee
and the constant $L_{0,3}$ can be taken $\frac{4}{\p^2 3^{\frac32}}$ as
in \cite[p.475]{Li984}, and $|\O|=L^3$ if $L$ is the container side.
\*

Since the Kolmogorov momentum scale $K_\n$ is proportional to
$K_\n=\frac1L\Big(\frac{\h}{\n^3}\Big)^{\frac14}$, \cite{Fr995},
this can be interpreted as saying that the number of degrees of
freedom resposible for the chaotic evolution $\lis\NN$ is of the
order of the number of harmonics with
momemtum below Kolmogorov's momentum scale (\ie with wave length
above Kolmogorov's length scale). 

The above statements hold for $INS^N$ independently of $N$; they
would also apply, with minor variations, to $RNS^N$ equations in
the (very unlikey) case that $\a(\uu_t)$ is eventually $\ge\e$
for some $\e>0$.

\def\SEC{RNS Lyapunov spectrum}
\section{\SEC}
\label{sec6}
\iniz

The Lyapunov exponents (LE) and the average spectrums (AS) of the
symmetrized linearization matrix are not averages of local
observables, so that the conjectures do not imply a relation
between such quantities under the equivalence condition.

Nevertheless asymptotic, as $N\to\infty$, equality to $\n$ of the
averages of $\a(\uu)$, Eq.\equ{e4.2}, in corresponding $INS^N$
and $RNS^N$, appears to hold in 2D simulations seems (at large
$N$): since $\a(\uu)$ is non local, if regarded as an observable,
this in $RNS^N$ can be shown to be a simple consequence of the
conjecture, \cite{Ga020b,Ga021}, {\it but not in
$INS^N$}.

Therefore it is natural to look if there are other non local
observables to which the equivalence can be extended: and results
in SM provide important examples of non local observables which
have equal average values in corresponding distributions of
different ensembles.

A {\it few} simulations have been performed on the NS problem in the
above context.  The results confirm the equality 
of the averaged Lyapunov spectra, \cite[Sec.18]{Ga020b},\cite{Ga021}.

In 2D a further property emerges from the simulations: if the $M$
average local LE's (defined inthe previous section) for $RNS^N$
are labeled $\l_k$ with $k\in [0,M)$, ordered by decreasing size,
and if $\overline{div}$ is the average phase space contraction
rate\footnote{\small Contraction rate = trace of the above matrix
$J(\uu,x)$.} an appoximate ``{\it pairing rule}'' appears:
\be \frac12(\l_k+\l_{M-k})=\frac{1}{M}\overline{div},\quad
k=0,\ldots, \frac{M}2-1\Eq{e6.1}\ee
The latter relation is well verified in the simulations with
small regularization $N$, so far performed, for all but a few
small $k$'s: and a question is whether the discrepansies remain
as $N\to\infty$. However (as hinted in \cite{Ga997b}) it is
expected that $(\l_k+\l_{M-k})$ for large $k$ (hence large $N$)
is a concave curve.

A pairing rule, rigorously holds in systems governed by a
Hamiltonian of the form $H=\frac12 \pp\cdot\pp$ and subject to a
force $\V f(\qq)$ only locally conservative and to the constraint
of maintaining constant kinetic energy $\frac12\pp^2$ via a force
$-\a(\pp,\qq) \pp$ (hence $\a=\frac{\pp\cdot\V
f(\qq)}{\pp\cdot\pp}$). Hence it is expected to hold for the
fluids decribed by adding a linear friction (``Ekmann friction'',
$-\n \uu$) to the Euler equations in Lagrangian form (\ie with a
second equation describing the individual fluid elements
trajectories, thus doubling the number of degrees of freedom and
of exponents): in \cite{Ga997b,Ga020b} the question is
contemplated about a possible connection between the above
pairings.
\*


{\bf Conflict of interest: \rm The author states that there is no
conflict of interest.}

\IfFileExists{./267.bbl}{\bibliography{267.bbl}}{\bibliography{0Bib}}
\end{document}